%% file: dual_pic.tex
\documentclass[notitlepage,showpacs,preprintnumbers,aps,pop,preprint]{revtex4-1}

\input{code_preamble}

\begin{document}

\title{Dual PIC: a structure preserving method for discretizing Lie-Poisson brackets}
\author{William Barham}
\email{william.barham@utexas.edu}
\affiliation{Oden Institute for Computational Engineering and Sciences, The University of Texas at Austin,  Austin, TX, 78712, USA}
\author{Philip J. Morrison}
\email{morrison@physics.utexas.edu}
\affiliation{Department of Physics and Institute for Fusion Studies, 
The University of Texas at Austin, Austin, TX, 78712, USA}
\date{\today}

\begin{abstract}
We consider a general discretization strategy for Hamiltonian field theories generated by Lie-Poisson brackets which we call dual PIC (DPIC). This method involves prescribing two different discrete representations of the dynamical variable which are constrained as a Casimir invariant of the flow to coincide with one another via an $L^2$ projection throughout the entire simulation. This allows one to leverage the relative advantages of each discrete representation. We begin by describing DPIC as applied to a general Lie-Poisson system and then provide illustrative examples:  the discretization of the two-dimensional vorticity equations and the Vlasov-Poisson equation. 
\end{abstract}

 \maketitle

\section{Introduction}

In this preliminary note we document a new computational method, which we call {\it dual PIC} or {\it DPIC} for short,  that brings together two main notions:  the particle-in-cell (PIC) method that was proposed early on in the plasma physics community (see e.g.\ \cite{hock,bird}) and the noncanonical Hamiltonian description possessed by dissipation free models in mechanics,  fluid mechanics, kinetic theory, plasma physics, and other fields of research (see \cite{pjm82,pjm98}). 

These two notions were brought together in GEMPIC \cite{GEMPIC}, which used the Vlasov-Maxwell Hamiltonian structure given in \cite{pjm80, pjm82, MW} to construct a Poisson integrator, an integrator that preserves Casimir invariants while being symplectic on their invariant sets. This was accomplished by merging the particle description of the Vlasov phase space density, which is naturally Hamiltonian, with finite element exterior calculus (FEEC) originated in \cite{arnold} for representation of the electromagnetic fields of Maxwell's equations. The net result of these discretizations is a finite system of ordinary differential equations that can be generated from a noncanonical Poisson bracket, i.e., a Poisson bracket of nonstandard form with degeneracy (see e.g.\ \cite{pjm98}). 

A special class of noncanonical Hamiltonian systems are known as Lie-Poisson systems. These systems possess a noncanonical Poisson brackets that are linear in the dynamical variables of the system and are in correspondence with  Lie algebras, finite or infinite. Unlike PIC methods that couple kinetic theories with electromagnetism, fluid particle methods do not so easily produce Hamiltonian discretizations. The method proposed here does this by adding a dual field to a Lie-Poisson Hamiltonian partial differential equation, which roughly speaking plays the role of the Maxwell fields.  In the context of the two-dimensional Euler fluid equation for vorticity, described here as an example,  the vorticity has two descriptions, one in terms of point vortices and another in terms of a Galerkin basis.  The two descriptions are then naturally seen to be tied together by a Casimir invariant of the discretized system. In the context of the Vlasov-Poisson equation, which likewise possesses Lie-Poisson structure \cite{pjm80}, the dual field provides a regularized representation of the phase-space density which in a PIC method is only defined distributionally. 

This note is organized as follows.  In Sec.\  \ref{liepoisson} we discuss general Lie-Poisson systems and our method of discretization. This is followed in Secs.\ \ref{vlasov} and \ref{euler} where application to the Vlasov-Poisson equation and two-dimensional Euler fluid equation are addressed, respectively.  
We conclude in Sec.\ \ref{conclu} where we sum up, discuss generalizations,  and future work. 

\section{The general dual PIC method} \label{liepoisson}
\input{sections/general_dpic}

\section{Application to the Vlasov-Poisson equation} \label{vlasov}
\input{sections/vlasov-poisson}

\section{Application to two-dimensional incompressible flows} \label{euler}
\input{sections/2d_euler}

\section{Conclusion} \label{conclu}
\input{sections/conclusion}

\section{Acknowledgements}
We gratefully acknowledge the support of U.S. Dept. of Energy Contract \# DE-FG05-80ET-53088, NSF Graduate Research Fellowship \# DGE-1610403, and the Humboldt foundation.


\bibliographystyle{apsrev}

\bibliography{references}


\appendix

\section{Appendix: proof of Jacobi identity}
\label{Ajacprf}
\input{sections/appendix}

\end{document}

%% file: code_preamble.tex
\usepackage{amsmath,amssymb,bm,amsthm,amsfonts,amstext,graphics,graphicx,subfigure}
\usepackage[colorlinks]{hyperref}
\usepackage{color}
\usepackage{mathbbol}
\usepackage{enumitem}
\usepackage{geometry}              
\usepackage{upgreek} 

\newtheorem{theorem}{Theorem}

\newtheorem{proposition}[theorem]{Proposition}

\def\XXint#1#2#3{{\setbox0=\hbox{$#1{#2#3}{\int}$ }
\vcenter{\hbox{$#2#3$ }}\kern-.5\wd0}}

\def\bq{\begin{equation}}
\def\eq{\end{equation}}
\def\bqy{\begin{eqnarray}}
\def\eqy{\end{eqnarray}}

\def\bal#1\eal{\begin{align}#1\end{align}}

\newcommand{\msf}[1]{\mathsf{#1}}
\newcommand{\bmsf}[1]{\boldsymbol{\mathsf{#1}}}

%% file: sections/general_dpic.tex
The discretization strategy proposed herein may be applied to canonical Lie-Poisson systems of the following form. Let $\mu$ be a scalar field with domain $T \mathbb{R}^n \sim \mathbb{R}^{2n}$. Moreover, suppose that $\mu$ evolves as a canonical Lie-Poisson system:
\begin{equation}
	\partial_t \mu = \{ \mu, H\}
\end{equation}
where
\begin{equation}
	\{F, G\} = \int_{T \mathbb{R}^n} \mu \left[ \frac{\delta F}{\delta \mu}, \frac{\delta G}{\delta \mu} \right] \mathsf{d}^{n} \bm{z} \quad \text{and} \quad H[\mu] = \int_{T \mathbb{R}^n} \mu \frac{\delta K}{\delta \mu} \mathsf{d}^{n} \bm{z}
\end{equation}
where $\bm{z} \in T \mathbb{R}^n$, $[f, g] = \nabla_{\bm{z}} f^T J_c \nabla_{\bm{z}} g$ is the finite dimensional canonical Poisson bracket, $\mu$ is interpreted as a density, and $K[\bm{z}, \mu]$ is interpreted as a kind of  energy functional. Systems of such form are relatively abundant in plasma physics with the Vlasov-Poisson system being a notable example \cite{morrison-1980_maxwell-vlasov}. 

The key observation underlying the dual PIC method is that it is possible to extend the system to possess two density variables which coincide with each other by design. To this end we extend the continuous system by defining a Lie-Poisson system with two density variables, $\mu_1$ and $\mu_2$:
\begin{equation}
	\{F, G\}_E \!=\! \frac{1}{2}\! \left(\! \mu_1, \left[ \frac{\delta F}{\delta \mu_1} + \frac{\delta F}{\delta \mu_2}, \frac{\delta G}{\delta \mu_1} + \frac{\delta G}{\delta \mu_2} \right] \right) 
	\   \text{and} \  \, 
	H_E[\mu_1, \mu_2] \!= \!\left(\!\mu_1, \frac{\delta K}{\delta \mu_2} (\bm{z}, \mu_2) \right)\!,
	\label{LPsystem}
\end{equation}
where $(\cdot, \cdot)$ is the $L^2$ inner product. The Poisson bracket of \eqref{LPsystem} for this extended system satisfies the Jacobi identity; we provide the proof in Appendix \ref{Ajacprf}.
Also, one may show that 
\begin{equation}
	C_1 = \mathcal{C}[\mu_1]
	\quad \text{and} \quad
	C_2 = \mathcal{C}[\mu_1 - \mu_2]
\end{equation}
are Casimir invariants for any functional $\mathcal{C}$. Hence, if $\left. \mu_1 \right|_{t = 0} = \left. \mu_2 \right|_{t = 0}$, then $\mu_1 = \mu_2 \ \forall t$. When $\mu_1 = \mu_2$, one recovers the original system. 

\subsection{Derivation of the general dual PIC method}
\label{dpicdisc}

We begin with the Lie-Poisson system of \eqref{LPsystem} and discretize.
 Let $\{ \psi_i \}_{i=1}^N$ be some Galerkin basis over $H^1(T \mathbb{R}^n)$. It is necessary that it be possible to take derivatives of the basis functions. Let $\{ \bm{z}_a \}_{a= 1}^{N_p} \subset T\mathbb{R}^n$ and $\{ w_a \}_{a = 1}^{N_p} \in \mathbb{R}$. We discretize by letting 
\begin{equation}
	\mu_1 = \sum_{a=1}^{N_p} w_a \delta(\bm{z}_a - \bm{z} ) \quad \text{and} \quad \mu_2 = \sum_{i=1}^N (\mu_*)_i \mathbb{M}_{ij}^{-1} \psi_j(\bm{z})\,,
\end{equation}
where $\mathbb{M}$ is the $L^2$ mass matrix associated with the Galerkin basis. The star subscript is a reminder that we have interpolated $\mu_2$ with respect to the dual Galerkin basis, $\{ \mathbb{M}_{ij}^{-1} \psi_j^0(\bm{x}) \}_{i=1}^{N}$. If we let $F[\mu_1, \mu_2] = \msf{F}(\bm{z}_1, \hdots, \bm{z}_{N_p}, \bm{\mu}_*)$, then one may show that
\begin{equation}
	\frac{\delta F}{\delta \mu_2} = \sum_{i=1}^N \frac{\partial \msf{F}}{\partial (\mu_*)_i} \psi_i(\bm{z})\,.
\end{equation}
If we interpolate $\mu_2$ with respect to the dual basis, functional derivatives with respect to $\mu_2$ are interpolated with respect to the primal basis. If we ever wish to evaluate $\mu_2$, it must be done using the dual basis. Therefore, following \cite{GEMPIC}, we find
\begin{equation}
	\left. \nabla_{\bm{z}} \frac{\delta F}{\delta \mu_1} \right|_{\bm{z} = \bm{z}_a} = \frac{1}{w_a} \frac{\partial \msf{F}}{\partial \bm{z}_a} 
	\quad \text{and} \quad 
	\left. \nabla_{\bm{z}} \frac{\delta F}{\delta \mu_1} \right|_{\bm{z} = \bm{z}_a} = \sum_{i=1}^N \frac{\partial \msf{F}}{\partial (\mu_*)_i} \nabla_{\bm{z}} \psi_i(\bm{z}_a).
\end{equation}
We may directly insert these discretized functional derivatives into the Lie-Poisson bracket and Hamiltonian. 

\subsection{Notation}

Before proceeding, we define a more compact notation. Let 
\begin{equation}
	\bm{Z} = 
	\begin{pmatrix}
		\bm{z}_1 \\
		\bm{z}_2 \\
		\vdots \\
		\bm{z}_{N_p}
	\end{pmatrix} \in \mathbb{R}^{2n N_p},
	\quad
	\bmsf{w} = 
	\begin{pmatrix}
		w_1 \\
		w_2 \\
		\vdots \\
		w_{N_p}
	\end{pmatrix},
	\quad
	\bm{\mu}_* =
	\begin{pmatrix}
		(\mu_*)_1 \\
		(\mu_*)_2 \\
		\vdots \\
		(\mu_*)_{N}
	\end{pmatrix},
	\quad
	\mathbb{M}_p = \text{diag}(\bmsf{w}) \otimes \mathbb{I}_{2n}\;,
\end{equation}	
\begin{equation}	
	\mathbb{J}_c = \mathbb{I}_{N_p} \otimes J_c,
	\quad \text{and} \quad
	\mathbb{\Psi}(\bm{Z}) = 
	\begin{pmatrix}
		\psi_1(\bm{z}_1) & \psi_2(\bm{z}_1) & \hdots & \psi_N(\bm{z}_1) \\
		\psi_1(\bm{z}_2) & \psi_2(\bm{z}_2) & \hdots & \psi_N(\bm{z}_2) \\
		\vdots & & \ddots & \vdots \\
		\psi_1(\bm{z}_{N_p}) & & & \psi_N(\bm{z}_{N_p})
	\end{pmatrix}
	\in \mathbb{R}^{N_p \times N}.
\end{equation}
The alternant matrix $\mathbb{\Psi}(\bm{Z})$ connects the Galerkin and particle representation. Its derivative is denoted by
\begin{equation}
	D \mathbb{\Psi}(\bm{Z}) = 
	\begin{pmatrix}
		\nabla_{\bm{z}} \psi_1(\bm{z}_1) & \nabla_{\bm{z}} \psi_2(\bm{z}_1) & \hdots & \nabla_{\bm{z}} \psi_N(\bm{z}_1) \\
		\nabla_{\bm{z}} \psi_1(\bm{z}_2) & \nabla_{\bm{z}} \psi_2(\bm{z}_2) & \hdots & \nabla_{\bm{z}} \psi_N(\bm{z}_2) \\
		\vdots & & \ddots & \vdots \\
		\nabla_{\bm{z}} \psi_1(\bm{z}_{N_p}) & & & \nabla_{\bm{z}} \psi_N(\bm{z}_{N_p})
	\end{pmatrix}
	\in \mathbb{R}^{2n N_p \times N}.
\end{equation}

\subsection{The fully discretized extended system}
Using this compact notation, one may write
\begin{eqnarray}
	\left( \left. \nabla_{\bm{z}} \left( \frac{\delta F}{\delta \mu_1} +  \frac{\delta F}{\delta \mu_2} \right) \right|_{\bm{z}} \right)_{a = 1, \hdots, N_p} 
	&=& \mathbb{M}_p^{-1} \frac{\partial \msf{F}}{\partial \bm{Z}} + \mathbb{D \Psi}(\bm{Z}) \frac{\partial \msf{F}}{\partial \bm{\mu}_*} 
	\nonumber\\
	&=& 
	\begin{pmatrix}
		\mathbb{M}_p^{-1} & D \mathbb{\Psi}(\bm{Z})
	\end{pmatrix}
	D \msf{F}(\bm{Z}, \bm{\mu}_*).
\end{eqnarray}
Letting $\{F, G\}_E = [ \msf{F}, \msf{G} ]_E$, one finds
\begin{equation} \label{eq:disc_poisson_bracket}
	\begin{aligned}
		\ [ \msf{F}, \msf{G} ]_E 
		&= 
		\frac{1}{2} D\msf{G}(\bm{Z}, \bm{\mu}_*)^T 
		\begin{pmatrix}
			\mathbb{M}_p^{-1} \\
			D \mathbb{\Psi}(\bm{Z})^T
		\end{pmatrix}
		\mathbb{M}_p \mathbb{J}_c
		\begin{pmatrix}
			\mathbb{M}_p^{-1} & D \mathbb{\Psi}(\bm{Z})
		\end{pmatrix}
		D \msf{F}(\bm{Z}, \bm{\mu}_*) \\
		&=:
		\frac{1}{2} D\msf{G}(\bm{Z}, \bm{\mu}_*)^T \mathbb{J}(\bm{Z}) D \msf{F}(\bm{Z}, \bm{\mu}_*).
	\end{aligned}
\end{equation}
The discretized Hamiltonian is given by
\begin{equation}
	H_E[\mu_1, \mu_2] = \msf{H}_E(\bm{Z}, \bm{\mu}_*) = \bmsf{w}^T \mathbb{\Psi}(\bm{Z}) \frac{\partial \msf{K}}{\partial \bm{\mu}}(\bm{Z}, \bm{\mu}_*).
\end{equation}
The dynamics are given by $\dot{\msf{F}} = [ \msf{F}, \msf{H}_E]_E$. 

\subsection{Restriction to a Casimir leaf}
A family of Casimir invariants may be found which satisfy:
\begin{equation}
	\begin{pmatrix}
		\mathbb{M}_p^{-1} & D \mathbb{\Psi}(\bm{Z})
	\end{pmatrix}
	D \msf{F}(\bm{Z}, \bm{\mu}_*) = 0 \iff \mathbb{M}_p^{-1} \frac{\partial \msf{F}}{\partial \bm{Z}} + D \mathbb{\Psi}(\bm{Z}) \frac{\partial \msf{F}}{\partial \bm{\mu}_*} = 0.
\end{equation}
In particular, we find that these equations are  satisfied for the family of Casimir invariants:
\begin{equation}
	C_i(\bm{\mu}, \bm{Z}) = (\mu_*)_i - \sum_{a=1}^{N_p} w_a \psi_i(\bm{z}_a) \quad \text{or} \quad \bm{C}(\bm{\mu}, \bm{Z}) = \bm{\mu}_* - \bmsf{w}^T \mathbb{\Psi}(\bm{Z}).
\end{equation}
Therefore, 
\begin{equation}
	\left. \mathbb{\Psi}(\bm{Z})^T \bmsf{w} \right|_{t = 0} = \left. \bm{\mu}_* \right|_{t = 0} \implies \mathbb{\Psi}(\bm{Z})^T \bmsf{w} = \bm{\mu}_* \ \forall t.
\end{equation}

If we restrict dynamics to live on the Casimir leaf $\bmsf{w}^T \mathbb{\Psi}(\bm{Z}) = \bm{\mu}_*$, we may eliminate $\bm{\mu}_*$ from our system. This yields
\begin{equation}
	[ \msf{F}, \msf{G}] = D\msf{F}(\bm{Z})^T \mathbb{M}_p^{-1} \mathbb{J}_c D\msf{G}(\bm{Z}) \quad \text{and} \quad \msf{H}(\bm{Z}) = \left. \bmsf{w}^T \mathbb{\Psi}(\bm{Z}) \frac{\partial \msf{K}}{\partial \bm{\mu}_*} \right|_{\bm{\mu}_* = \mathbb{\Psi}(\bm{Z})^T \bmsf{w}}.
\end{equation}
Hence, we obtain the Hamiltonian system $\dot{\bm{Z}} = \mathbb{M}_p^{-1} \mathbb{J}_c D \msf{H}(\bm{Z})$. Notice that this fully discretized system possesses the same Poisson bracket as if we had discretized the original Lie-Poisson system via a particle method alone. The difference is that the Hamiltonian includes interpolation in a Galerkin basis. 

At any point in the simulation, we may reconstruct the coefficients in the Galerkin basis via $\bm{\mu}_* = \bmsf{w}^T \mathbb{\Psi}(\bm{Z})$. We may then interpolate the field via
\begin{equation}
	\mu_h(\bm{z}) = \sum_{i=1}^{N} (\mu_*)_i \mathbb{M}^{-1}_{ij} \psi_j(\bm{z})\,.
\end{equation}
Given initial data prescribed by the function $\mu_0(\bm{z})$, if one draws $\left. \bm{Z} \right|_{t=0}$ randomly and uniformly from $T \mathbb{R}^n$, and computes $\left. \mu_i \right|_{t=0} = ( \psi_i, \mu_0 )$ for $i=1, \hdots, N$, then one may select the initial weights by finding the least squares solution of $\bm{\mu}_* = \bmsf{w}^T \mathbb{\Psi}(\bm{Z})$.

%% file: sections/vlasov-poisson.tex
The (nondimensional) Vlasov-Poisson equations are given by
\begin{equation}
	\begin{aligned}
		\partial_t f + \bm{v} \cdot \nabla_{\bm{x}} f - \nabla_{\bm{x}} \phi \cdot \nabla_{\bm{v}} f &= 0 \\
		- \Delta_x \phi = \int f(\bm{x}, \bm{v}, t) \mathsf{d}^3 \bm{v}\,,
	\end{aligned}
\end{equation}
where $(\bm{x}, \bm{v}) \in T\Omega \subset \mathbb{R}^3 \times \mathbb{R}^3$. As shown in \cite{morrison-1980_maxwell-vlasov}, these equations are generated by the Lie-Poisson bracket and Hamiltonian
\begin{equation*}
	\{F, G\} = - \int\! f \left[ \frac{\delta F}{\delta f}, \frac{\delta G}{\delta f} \right]\,  \mathsf{d}^3 \bm{x} \mathsf{d}^3 \bm{v} \ \  \text{and} \ \  H[f] = \frac{1}{2} \int \left(\! | \bm{v} |^2 + \phi(\bm{x}) \right) f(\bm{x}, \bm{v}, t)\,  \mathsf{d}^3 \bm{x} \mathsf{d}^3 \bm{v}\,.
\end{equation*}

\subsection{Derivation of dual PIC for Vlasov-Poisson}

Following the general development of Sec.~\ref{dpicdisc}, we consider two discrete representations of the phase-space density:
\begin{equation}
	f_1 = \sum_{a = 1}^{N_p} w_a \delta(\bm{x} - \bm{x}_a) \delta(\bm{v} - \bm{v}_a) \quad \text{and} \quad f_2 = \sum_{i,j = 1}^{N_x, N_v} \mathsf{f}_{ij} \psi^x_i(\bm{x}) \psi^v_j(\bm{v})\,.
\end{equation}
where $\{ \psi_i^x \}_{i=1}^{N_x}$ and $\{ \psi_i^v \}_{i=1}^{N_v}$ are Galerkin bases. Hence, we have decomposed our Galerkin basis over phase-space using tensor product shape functions. As in the general theory, we rather consider the Galerkin coefficients with respect to the dual basis:
\begin{equation}
	(\mathsf{f}_*)_{ij} = \sum_{i,j=1}^{N_x,N_v} \mathbb{M}_{ik}^x \mathbb{M}_{jl}^v \mathsf{f}_{kl}\,.
\end{equation}
From the general theory, we know that the $L^2$ projection constraining these two fields to coincide is given by
\begin{equation}
	(\mathsf{f}_*)_{ij} = \sum_{a=1}^{N_p} w_a \psi^x_i(\bm{x}_a) \psi^v_j(\bm{v}_a)\,.
\end{equation}
We write the extended continuous Hamiltonian as
\begin{equation}
	H_E[f_1, f_2] = \int \left( \frac{1}{2} | \bm{v} |^2 + \phi_2(\bm{x}) \right) f_1(\bm{x}, \bm{v}, t) \mathsf{d}^3 \bm{x} \mathsf{d}^3 \bm{v}\,,
\end{equation}
with
\begin{equation}
	- \Delta_{\bm{x}} \phi_2 = \int f_2(\bm{x}, \bm{v}) \mathsf{d}^3 \bm{v}\,.
\end{equation}
Applying our Galerkin representation of $f_2$, we find
\begin{equation}
	\Delta_{\bm{x}} \phi_2(\bm{x}) 
	= \left( \sum_{i=1}^{N_x} \left( \sum_{j=1}^{N_v} \mathsf{f}_{ij} \int \psi^v_j(\bm{v}) \mathsf{d}^3 \bm{v} \right) \psi^x_i (\bm{x}) \right) 
	\implies \upphi_i = \sum_{j=1}^{N_x} \mathbb{L}^{-1}_{ij} \rho_j \,,
\end{equation}
where $\rho_i =  \sum_{jk}  \mathbb{M}_{ik}^x \mathsf{f}_{kj} \int \psi^v_j(\bm{v}) \mathsf{d}^3 \bm{v}$ is the discrete charge density, $\mathbb{L}_{ij} = (\nabla \psi^x_i, \nabla \psi^x_j)$ is the discrete Laplacian in the Galerkin basis $\{ \psi^x_i \}$, and $\bm{\upphi}$ is the vector of coefficients for $\phi_2$ in the Galerkin basis. The discrete Hamiltonian is thus given by
\begin{equation}
	\begin{aligned}
		H_E[f_1, f_2] 
			&= \int \left( \frac{1}{2} | \bm{v} |^2 + \phi_2(\bm{x}) \right) f_1(\bm{x}, \bm{v}, t) \mathsf{d}^3 \bm{x} \mathsf{d}^3 \bm{v} \\
			&= \sum_{a=1}^{N_p} w_a \left( \frac{1}{2} | \bm{v}_a |^2 + \sum_{i, j=1}^{N_x} \mathbb{L}^{-1}_{ij} \rho_j \psi^x_i (\bm{x}_a) \right) \\
			&= \frac{1}{2} \bm{V}^T \mathbb{M}_p \bm{V} + \bmsf{w}^T \mathbb{\Psi}(\bm{X}) \mathbb{L}^{-1}
			 \bm{\rho}\,,
	\end{aligned}
\end{equation}
where $\mathbb{\Psi}(\bm{X}) \in \mathbb{R}^{N_p \times N}$ such that $[\mathbb{\Psi}(\bm{X})]_{ai} = \psi^x_i(\bm{x}_a)$. Using the $L^2$ projection to eliminate the Galerkin coefficients, this discrete Hamiltonian may be written entirely in terms of $(\bm{X}, \bm{V})$. From the general theory, we know that the dynamics are generated by the Poisson bracket
\begin{equation}
	[ \mathsf{F}, \mathsf{G} ] = 	
	\begin{pmatrix}
		0 & \mathbb{M}_p^{-1} \\
		- \mathbb{M}_p^{-1} & 0
	\end{pmatrix}.
\end{equation}

\subsection{Simplification using a partition of unity basis}

In principle, the method discussed above is a fully discrete  Hamiltonian theory. However, this approach suffers from the fact that the discrete charge $\bm{\rho}$, is velocity dependent through the Galerkin projection:
\begin{eqnarray}
	\rho_i &=& \sum_{j, k = 1}^{N_v} (\mathsf{f}_*)_{ij} (\mathbb{M}^v)^{-1}_{jk} \int \psi^v_k(\bm{v}) \mathsf{d}^3 \bm{v} 
	\nonumber\\
	&=& \sum_{a=1}^{N_p} \sum_{j, k = 1}^{N_v} \left( w_a (\mathbb{M}^v)^{-1}_{jk} \int \psi^v_k(\bm{v}) \mathsf{d}^3 \bm{v} \right) \psi^x_i(\bm{x}_a) \psi^v_j(\bm{v}_a)\,.
\end{eqnarray}
This prevents us from using Hamiltonian splitting in our time-stepping scheme. However, this may be remedied using a partition of unity basis in velocity space.
\begin{proposition}
\begin{equation}
	\sum_{j=1}^{N_v} \psi^v_j(\bm{v}) = 1 \quad \forall \bm{v}
	\implies
	\sum_{k=1}^{N_v} (\mathbb{M}^v)^{-1}_{jk} \int \psi^v_k(\bm{v}) \mathsf{d}^3 \bm{v} = 1 \quad \forall j.
\end{equation}
\end{proposition}
\noindent \textbf{Proof:} Define $\mathbb{1}_i = 1$ $\forall i$. Then
\begin{eqnarray}
		&& \sum_{j=1}^{N_v} \psi^v_j(\bm{v}) = 1 \implies \sum_{k=1}^{N_v} \left( \psi^v_j, \psi^v_k \right) = \int \psi^v_j(\bm{v}) \mathsf{d}^3 \bm{v}
		\nonumber
	\\
	&& \implies (\mathbb{M}^v \mathbb{1})_j = \int \psi^v_j(\bm{v}) \mathsf{d}^3 \bm{v}\,. \qed
\end{eqnarray}
Such a basis may be constructed using, for example, B-splines. Hence, in this case, we find that
\begin{equation}
	\rho_i = \sum_{a=1}^{N_p}  w_a \psi^x_i(\bm{x}_a) \sum_{j=1}^{N_v} \psi^v_j(\bm{v}_a) 
	= \sum_{a=1}^{N_p} w_a \psi^x_i(\bm{x}_a)
	= (\mathbb{\Psi}(\bm{X})^T \bmsf{w})_i\,.
\end{equation}
Hence, we find the Hamiltonian reduces to
\begin{equation}
	\mathsf{H}(\bm{X}, \bm{V}) = \frac{1}{2} \bm{V}^T \mathbb{M}_p \bm{V} + \bmsf{w}^T \mathbb{\Psi}(\bm{X}) \mathbb{L}^{-1} \mathbb{\Psi}(\bm{X})^T \bmsf{w}\,,
\end{equation}
and the equations of motion are given by
\begin{equation}
	\dot{\bm{X}} = \bm{V}
	\quad \text{and} \quad
	\dot{\bm{V}} = - D \mathbb{\Psi}(\bm{X}) \mathbb{L}^{-1} \mathbb{\Psi}(\bm{X})^T \bmsf{w}\,,
\end{equation}
where $D \mathbb{\Psi}(\bm{X}) = \partial \mathbb{\Psi} / \partial \bm{X}$. 

\subsection{One-dimensional Vlasov-Poisson algorithm}

We now present the algorithm applied to the 1D Vlasov-Poisson system in full. Suppose $(x,v) \in \mathbb{T}^2$ so that we are solving the 1D Vlasov-Poisson system. Moreover, suppose that we choose a partition of unity basis $\{ \psi_i^v \}_{i=1}^{N_v}$ in $v$ and a general Galerkin basis $\{ \psi_i^x \}_{i=1}^{N_x}$ in $x$. The dynamical variables are the particle positions: $(\bm{X}, \bm{V}) = ( [x_a]_{a=1}^{N_p}, [v_a]_{a=1}^{N_p}) \subset \mathbb{T}^2$. We find that the equations of motion are
\begin{equation}
	\dot{\bm{X}} = \bm{V}
	\quad \text{and} \quad
	\dot{\bm{V}} = - D \mathbb{\Psi}(\bm{X}) \mathbb{L}^{-1} \mathbb{\Psi}(\bm{X})^T \bmsf{w}\,.
\end{equation}
To solve this problem numerically, the first step is to find $\bmsf{w}$. We find that
\begin{equation}
	\sum_{a=1}^{N_p} w_a \psi^x_i(x_a) = \sum_{j=1}^{N_v} (\mathsf{f}^0_*)_{ij} = \sum_{j=1}^{N_v} \int_{\mathbb{T}^2} f_0(x, v) \psi^x_i(x) \psi^v_j(v) \mathsf{d} x \mathsf{d} v\,.
\end{equation}
where $\mathbb{M}^x$ and $\mathbb{M}^v$ are the mass matrices for the $x$ and $v$ Galerkin bases. We break the selection of the initial weights into three steps:
\begin{enumerate}
	\item Form the matrices $(\msf{f}_*^0)_{kl} = \int_{\mathbb{T}^2} f_0(x, v) \psi^x_k(x) \psi^v_l(v) \mathsf{d} x \mathsf{d} v$, $\mathbb{M}^x$, and $\mathbb{M}^v$,
	\item Uniformly draw $(\bm{X}^0, \bm{V}^0)$ from $\mathbb{T}^2$,
	\item Find the least squares solution of $\mathbb{\Psi}(\bm{X}^0)^T \bmsf{w} = \msf{f}^0_* \mathbb{1}$ where $\mathbb{1}_i = 1$ $\forall i$.
\end{enumerate}
The first step is particularly easy if the initial distribution is a tensor product, $f_0(x, v) = f_0^x(x) f_0^v(v)$.

Once we have the initial weights, time evolution is accomplished by Hamiltonian splitting:
\begin{equation*}
	\begin{split}
		\dot{\bm{X}} &= \bm{V} \\
		\dot{\bm{V}} &= 0
	\end{split}
	\qquad \text{and} \qquad
		\begin{split}
		\dot{\bm{X}} &= 0 \\
		\dot{\bm{V}} &= - D \mathbb{\Psi}(\bm{X}) \mathbb{L}^{-1} \mathbb{\Psi}(\bm{X})^T \bmsf{w}\,.
	\end{split}
\end{equation*}
We then compose these two flows together, for example via second order Strang splitting, to evolve the system:
\begin{equation}
	\begin{aligned}
		\bm{X}^{n+1/2} &= \frac{\Delta t}{2} \bm{V}^n \\
		\bm{V}^{n+1} &= - \Delta t D \mathbb{\Psi}(\bm{X}^{n+1/2}) \mathbb{L}^{-1} \mathbb{\Psi}(\bm{X}^{n+1/2})^T \bmsf{w} \\
		\bm{X}^{n+1} &= \frac{\Delta t}{2} \bm{V}^{n+1}\,.
	\end{aligned}
\end{equation}
At any given time step, we may reconstruct the solution in the Galerkin basis via:
\begin{align}
	\sum_{k,l=1}^{N_x,N_v} \mathbb{M}_{ik}^x \mathbb{M}_{jl}^v \msf{f}_{kl}^n &= (\mathsf{f}_*)_{ij}^{n} = \sum_a w_a \psi^x_i(x_a^n) \psi^v_j(v_a^n)
	\nonumber\\
	& \implies f_h^n(x,v) = \sum_{i,j=1}^{N_x,N_v} \msf{f}_{ij}^{n} \psi^x_i(x) \psi^v_j(v)\,.
\end{align}

%% file: sections/2d_euler.tex
Let $\Omega \subset \mathbb{R}^2$. The scalar vorticity, $\omega$, in two-dimensions evolves according to
\begin{equation}
	\partial_t \omega = - \nabla^\perp \phi \cdot \nabla \omega, \quad \text{where} \quad \Delta \phi = \omega
\end{equation}
and $\nabla^\perp \phi = (\partial_y \phi, - \partial_x \phi)$. Assuming all boundary terms may be neglected, these dynamics are generated by the following Poisson bracket and Hamiltonian \cite{pjm98}:
\begin{equation}
	 \{F, G\} = \int_\Omega \omega \left[ \frac{\delta F}{\delta \omega}, \frac{\delta G}{\delta \omega} \right] \mathsf{d}^2 \bm{x}\quad \mathrm{and}\quad H[\omega] = - \frac{1}{2} \int_\Omega \omega \phi \,  \mathsf{d}^2 \bm{x}\,,
\end{equation}
where $[f,g] = f_x g_y - f_y g_x$.

We represent two different vorticity fields by $\omega_1, \omega_2$, again with a Poisson bracket of the form 
\begin{equation}
	\{F, G\} = \frac{1}{2} \int_\Omega \omega_1 \left[ \frac{\delta F}{\delta \omega_1} + \frac{\delta F}{\delta \omega_2}, \frac{\delta G}{\delta \omega_1} + \frac{\delta G}{\delta \omega_2} \right] \mathsf{d}^2 \bm{x}\,,
\end{equation}
with  Casimir invariants
\begin{equation}
	C_1 = \mathcal{C}_1[\omega_1] 
	\quad \text{and} \quad 
	C_2 =\mathcal{C}_2[\omega_1 - \omega_2]\,.
\end{equation}
Consider now the Hamiltonian
\begin{equation}
	H[\omega_1, \omega_2] = - \int_\Omega \omega_1 \phi_2 \,  \mathsf{d}^2 \bm{x}
\end{equation}
where $\Delta \phi_2 = \omega_2$. Hence,
\begin{equation}
	\frac{\delta H}{\delta \omega_1} = - \phi_2 \quad \text{and} \quad \frac{\delta H}{\delta \omega_2} = - \phi_1 \implies \frac{\delta H}{\delta (\omega_1 + \omega_2)} = - \left( \phi_1 + \phi_2 \right).
\end{equation}
This yields the system of equations
\begin{eqnarray}
	&&\dot{F}[ \omega_1, \omega_2] = \{F, H \} = \frac{1}{2} \int_\Omega \left( \frac{\delta F}{\delta \omega_1} + \frac{\delta F}{\delta \omega_2} \right) \big[ \omega_1, \left( \phi_1 + \phi_2 \right) \big] \mathsf{d}^2 \bm{x}
	\nonumber\\
	&&  \implies \quad
	\begin{aligned}
		\partial_t \omega_1 &= \frac{1}{2} [ \omega_1, \phi_1 + \phi_2] \\
		\partial_t \omega_2 &= \frac{1}{2} [ \omega_1, \phi_1 + \phi_2].
	\end{aligned}
\end{eqnarray}
If $\omega_1 = \omega_2$, this clearly reduces to the usual vorticity equation. Note that Casimir invariant $C_2$ ensures that
\begin{equation}
	\left. \omega_1 \right|_{t=0} = \left. \omega_2 \right|_{t=0} \implies \omega_1 = \omega_2 \quad \forall t.
\end{equation}
Hence, standard 2D vortex dynamics lives on a Casimir leaf of the extended system. 

\subsection{DPIC Discretization}
DPIC discretization of the two-dimensional vorticity equations is given by
\begin{equation}
	\omega_1 = \sum_{a=1}^{N_p} w_a \delta(\bm{x} - \bm{x}_a(t))
	\quad \text{and} \quad
	\omega_2 = \sum_{i,j=1}^{N} (\upomega_*)_i \mathbb{M}^{-1}_{ij} \psi_j(\bm{x}).
\end{equation}
As with the general DPIC method,
\begin{equation}
	\frac{\delta F}{\delta \omega_2} = \sum_i \frac{\partial F}{\partial (\upomega_*)_i} \psi_j(\bm{x}) 
	\quad \text{and} \quad 
	\frac{\partial F}{\partial \bm{x}_a} = \left. w_a \nabla \frac{\delta F}{\delta \omega_1} \right|_{\bm{x}_a}\,,
\end{equation}
where $\mathbb{M}$ is the finite element mass matrix associated with the Galerkin basis. We use the same compact notation as in the previous sections and find that the discretized Poisson bracket is the same as in the general case (see \eqref{eq:disc_poisson_bracket}).

If we used a Hamiltonian which makes reference to the stream function associated with $\omega_1$, then we would have to solve
\begin{equation}
	\Delta \phi_1 = \sum_{a=1}^{N_p} w_a \delta(\bm{x} - \bm{x}_a).
\end{equation}
Hence, $\phi_1$ is expressed with the Green's function for the Laplacian. In this case, the Green's function and Hamiltonian are given by
\begin{equation} \label{eq:greens_func}
	G(\bm{x}, \bm{y}) = \frac{1}{2 \pi} \log(| \bm{x} - \bm{y}|) \implies H[\omega_1] = - \frac{1}{4 \pi} \sum_{\substack{ a, b = 1 \\ a \neq b}} w_a w_b \log( | \bm{x}_a - \bm{x}_b | ).
\end{equation}
Due to the non-compact support of the Green's function, each particle is advected by every other particle. This is a well-known issue for point vortex dynamics that may be circumvented using fast multipole methods \cite{10.1016/0021-9991(87)90140-9}. However, we employ a different strategy. Instead, we only ever use the stream function associated with $\omega_2$.

Assuming homogeneous or periodic boundary conditions, to find $\phi_2$, we solve
\begin{equation}
	-( \nabla \psi, \nabla \phi_2) = (\psi, \omega_2) \quad \forall \psi \in \psi(\Omega) \implies \bm{\upphi} = \mathbb{L}^{-1} \bm{\upomega}_*\,,
\end{equation}
where $\mathbb{L}$ is the discrete Laplacian in our Galerkin basis. This yields the Hamiltonian 
\begin{equation}
	H(\bm{X}, \bm{\upomega}) 
		= - \sum_{a=1}^{N_p} \sum_{i = 1}^N w_a \upphi_i \psi_i(\bm{x}_a) = - \bm{\upphi}^T \mathbb{\Psi}(\bm{X})^T \bmsf{w}
		= - \bm{\upomega}_*^T \mathbb{L}^{-1} \mathbb{\Psi}(\bm{X})^T \bmsf{w} \,.
\end{equation}
Hence, using the fact that
\begin{equation}
	\frac{\partial ( \bmsf{w}^T \mathbb{\Psi}(\bm{X}))}{\partial \bm{X}} =  \mathbb{M}_p D \mathbb{\Psi}(\bm{X}) \in \mathbb{R}^{2 N_p \times N}\,,
\end{equation}
we find
\begin{equation}
	\frac{\partial H}{\partial \bm{X}} = - \mathbb{M}_p D \mathbb{\Psi}(\bm{X}) \mathbb{L}^{-1} \bm{\upomega}_*
	\quad \text{and} \quad
	\frac{\partial H}{\partial \bm{\upomega}_*} = - \mathbb{L}^{-1} \mathbb{\Psi}(\bm{X})^T \bmsf{w}.
\end{equation}

The equations of motion may be found using the fact that 
\begin{equation}
\dot{F}(\bm{X}, \bm{\upomega}_*) = \mathbb{J}(\bm{X}) DH(\bm{X}, \bm{\upomega}_*)\,.
\end{equation}
 To this end, we find that
\begin{equation}
	\begin{pmatrix}
		\mathbb{M}_p^{-1} & D \mathbb{\Psi}(\bm{X}) 
	\end{pmatrix}
	DH(\bm{X}, \bm{\upomega}_*)
	= - D \mathbb{\Psi}(\bm{X}) \mathbb{L}^{-1} \mathbb{\Psi}(\bm{X})^T \bmsf{w} - D \mathbb{\Psi}(\bm{X}) \mathbb{L}^{-1} \bm{\upomega}_*.
\end{equation}
Hence, we find that
\begin{equation}
	\begin{aligned}
		\dot{\bm{X}} &= - \mathbb{J}_c \left[ D \mathbb{\Psi}(\bm{X}) \mathbb{L}^{-1} \mathbb{\Psi}(\bm{X})^T \bmsf{w} 
					+ D \mathbb{\Psi}(\bm{X}) \mathbb{L}^{-1} \bm{\upomega}_* \right] \\
		\dot{\bm{\upomega}}_* &= - D \mathbb{\Psi}(\bm{X})^T \mathbb{M}_p \mathbb{J}_c \left[ D \mathbb{\Psi}(\bm{X}) \mathbb{L}^{-1} \mathbb{\Psi}(\bm{X})^T \bmsf{w} 
					+ D \mathbb{\Psi}(\bm{X}) \mathbb{L}^{-1} \bm{\upomega}_* \right] \\
					&=  D \mathbb{\Psi}(\bm{X})^T \mathbb{M}_p \dot{\bm{X}}\,.
	\end{aligned}
\end{equation}
We see that the equations of motion for the finite element coefficients may be expressed in terms of the evolution of the particle positions. This points to the fact that the we may eliminate the Galerkin variable from our theory. Just as in the continuous theory, this may be accomplished by considering a Casimir invariant of the bracket. 

As noted in the section on general Galerkin bases, the family of Casimir invariants
\begin{equation}
	\bm{C}(\bm{X}, \bm{\upomega}_*) = \mathbb{\Psi}(\bm{X})^T \bmsf{w} - \bm{\upomega}_*
\end{equation}
is conserved by the dynamics. If we constrain ourselves to live on the Casimir leaf where $\mathbb{\Psi}(\bm{X})^T \bmsf{w} = \bm{\upomega}_*$, then we obtain the following equation of motion:
\begin{equation}
	\dot{\bm{X}} = - \mathbb{J}_c D \mathbb{\Psi}(\bm{X}) \mathbb{L}^{-1} \mathbb{\Psi}(\bm{X})^T \bmsf{w}\,.
\end{equation}
Hence, we have obtained a particle based discretization which computes the stream function by taking a detour through a Galerkin basis. The Poisson bracket and Hamiltonian  are given by
\begin{equation}
	 \{F,G\} = DF(\bm{X})^T \mathbb{M}_p^{-1} \mathbb{J}_c DG(\bm{X})
	 \quad\mathrm{and}\quad
	 H(\bm{X}) = - \bmsf{w}^T \mathbb{\Psi}(\bm{X}) \mathbb{L}^{-1} \mathbb{\Psi}(\bm{X})^T \bmsf{w}\,.
\end{equation}

\subsection{Discretization with Truncated Fourier Basis}
Now let $\Omega = [0, 2 \pi]^2$. If we use a Fourier representation for the Galerkin basis, $\psi_{\bm{k}} (\bm{x}) = e^{i \bm{k} \cdot \bm{x}}/\sqrt{2 \pi}$, then the Galerkin representation is given by a truncated Fourier series:
\begin{equation}
	\omega_h(\bm{x}) = \frac{1}{\sqrt{2 \pi}} \sum_{\| \bm{k} \|_{\infty} = 0}^N \upomega_{\bm{k}} e^{i \bm{k} \cdot \bm{x}}\,,
\end{equation}
where $\| \bm{k} \|_{\infty} = \max \{ |k_1|, |k_2| \}$. Hence, in this truncated Fourier basis, 
\begin{equation}
	\Delta \phi = \omega \implies \upphi_{\bm{k}} = -\| \bm{k} \|_2^{-2} \upomega_{\bm{k}} \quad \text{for } \bm{k} \neq 0
\end{equation}
and $\upphi_0 = 0$ if we assume periodic boundary conditions and zero average. As $\phi$ is only uniquely specified up to a constant, we are free to make this assumption. Hence,
\begin{equation}
	H(\bm{\upomega}) = - \int_{[0,2 \pi]^2} \phi \omega\,  \mathsf{d}^2 \bm{x} = \sum_{\| \bm{k} \|_\infty = 1}^{N} \| \bm{k} \|_2^{-2} \upomega_{\bm{k}} \upomega_{-\bm{k}}\,.
\end{equation}
Now, if we use the fact that
\begin{equation}
	\upomega_{\bm{k}} = \frac{1}{\sqrt{2 \pi}} \sum_{a = 1}^{N_p} \mathsf{w}_a e^{i \bm{k} \cdot \bm{x}_a}\,,
\end{equation}
we obtain the Hamiltonian
\begin{equation}
	H(\bm{X}) = \frac{1}{\pi} \sum_{\| \bm{k} \|_\infty = 1}^N \sum_{ \substack{a, b = 1 \\ a \neq b}}^{N_p} \frac{\mathsf{w}_a \mathsf{w}_b \cos[ \bm{k} \cdot( \bm{x}_a - \bm{x}_b )]}{\| \bm{k} \|_2^2}\,,
\end{equation}
where we have omitted all constant terms in the Hamiltonian. We can see that this is an approximation of the Green's function in  \eqref{eq:greens_func} in the truncated Fourier basis evaluated at the particle positions. Hence, we recover point vortex dynamics in the limit as $N \to \infty$ as one might hope. While the truncated Fourier basis aids in understanding the nature of the discretization procedure introduced in this article, it provides no advantage over simply using the Green's function as in traditional point vortex methods. A finite element basis would provide greater benefit because of its compact support.

%% file: sections/conclusion.tex
The DPIC method is a generalization of a standard particle based discretization with an advected Galerkin representation acting as a regularized alias for the particle based representation. Hence, for our examples, it acts as an extension of the PIC method for the Vlasov equation and as an extension of the point-vortex method for the incompressible two-dimensional Euler equation. In DPIC, the discretized Poisson bracket is identical to that of a particle based method while the Hamiltonian is typically modified to leverage the advantages of the Galerkin representation. Moreover, the Galerkin representation, being entirely determined by the flow of the particles, may be reconstructed on the fly as needed. Using a compactly supported Galerkin basis allows one to use local solvers of elliptic constraints (e.g. computing the stream function when solving the vorticity equation or the charge when solving the Vlasov-Poisson equation). 

In the context of the Vlasov-Poisson system, the Galerkin representation might be a useful tool in designing discrete metriplectic collision operators (see \cite{hirvijoki_and_adams-2017_landau-collision, hirvijoki_et_al-2018_metriplectic} for discretizations of metriplectic collision operators, and see \cite{morrison-1986_metriplectic} for the origin of the concept of metriplectic dynamics). High performance implementations of the same $L^2$ projection relating the particle and Galerkin representations of the phase-space density have been studied in \cite{https://doi.org/10.48550/arxiv.2205.06402}. In the context of the two-dimensional Euler equations, we saw using the truncated Fourier basis, that the DPIC method yields a method which is closely related to a point vortex method. Hence, the method is more likely to provide an advantage when a compactly supported Galerkin basis, such as a finite element basis, is used. Finally, this method might be used to discretize other related two-dimensional flows such as the quasi-geostrophic potential vorticity equations and the Hasegawa-Mima equations \cite{doi:10.1063/1.3194275}. 

%% file: sections/appendix.tex
In this section, we demonstrate some of the properties of the extended Poisson bracket used in the dual PIC method. Namely, that the Jacobi identity holds for the continuous and discretized brackets.

\noindent{\bf Continuous bracket:}

We begin by considering the continuous bracket:
\begin{equation}
	\{F, G\} = \int \mu_1 \left[ F_1 + F_2, G_1 + G_2 \right] \,\mathsf{d}^{n} \bm{z}\,,
\end{equation}
where subscripts indicate differentiation with respect to the corresponding variable; i.e,,  $F_1 = \delta F/\delta \mu_1$ and $F_2 = \delta F/\delta \mu_2$. Hence,
\begin{equation}
	\{F,G\}_1 = \frac{\delta \{F, G \}}{\delta \mu_1} = [F_1 + F_2, G_1 + G_2] 
	\quad \text{and} \quad 
	\{F,G\}_2 = \frac{\delta \{F, G \}}{\delta \mu_2} = 0\,,
	\label{Fdivs}
\end{equation}
where in \eqref{Fdivs} the equality is modulo second functional derivative terms, since by a general theorem \cite{pjm82} these terms all cancel.
Therefore,
\begin{eqnarray}
	\{ \{F, G\}, H\} &=& \int \mu_1 \big( \left[ \{ F, G \}_1 + \{F, G\}_2, H_1 + H_2 \right] \big)\, \mathsf{d}^{n} \bm{z}
	\nonumber\\
	& =& \int \mu_1 \big[ \left[ F_1 + F_2, G_1 + G_2 \right], H_1 + H_2 \big] \, \mathsf{d}^{n} \bm{z}\,. 
\end{eqnarray}
Hence, letting $f = F_1 + F_2$, $g = G_1 + G_2$, and $h = H_1 + H_2$, we find
\begin{equation}
	\{ \{F, G\}, H\} + \circlearrowleft_{F,G,H} = \int \mu_1 \big[[f,g],h\big] \,  \mathsf{d}^{n} \bm{z}+ \circlearrowleft_{f,g,h} = 0.
\end{equation}
Hence, the Jacobi identity holds for the continuous system. 

Briefly, we also discuss the Casimir invariants of the continuous bracket. A Casimir invariant $C$ will satisfy $[\mu_1, C_1 + C_2] = 0$. Hence, we immediately see that
\begin{equation}
	C^{(1)} = \int \mathcal{C}(\omega_1)\,  \mathsf{d}^{n} \bm{z}
	\quad \text{and} \quad 
	C^{(2)} = \int \mathcal{C}(\omega_1 - \omega_2)\,  \mathsf{d}^{n} \bm{z}
\end{equation}
are Casimir invariants for arbitrary $\mathcal{C}$ since
\begin{equation}
	\begin{split}
		C^{(1)}_1 &= \mathcal{C}'(\omega_1) \\
		C^{(1)}_2 &= 0
	\end{split}
	\qquad \text{and} \qquad
	\begin{split}
		C^{(2)}_1 &= \mathcal{C}'(\omega_1 - \omega_2) \\
		C^{(2)}_2 &= - \mathcal{C}'(\omega_1 - \omega_2).
	\end{split}
\end{equation}

\noindent{\bf Discrete bracket:}

The discrete bracket is most easily expressed as a block matrix:
\begin{eqnarray}
	\mathbb{J}(\bm{Z}) 
	&=& \begin{pmatrix}
			\mathbb{M}_p^{-1} \\
			D \mathbb{\Psi}(\bm{Z})^T
		\end{pmatrix}
		\mathbb{M}_p \mathbb{J}_c
		\begin{pmatrix}
			\mathbb{M}_p^{-1} & D \mathbb{\Psi}(\bm{Z})
		\end{pmatrix} 
		\nonumber\\
		&=&
		\begin{pmatrix}
			\mathbb{M}_p^{-1} \mathbb{J}_c & \mathbb{J}_c D \mathbb{\Psi}(\bm{Z}) \\
			D \mathbb{\Psi}(\bm{Z})^T \mathbb{J}_c & D \mathbb{\Psi}(\bm{Z})^T \mathbb{M}_p \mathbb{J}_c D \mathbb{\Psi}(\bm{Z})
		\end{pmatrix}.
\end{eqnarray}
Notice, the Poisson matrix only depends on $\bm{Z}$. Letting $\bm{\zeta} = (\bm{Z}, \bm{\mu})$, we know that the Jacobi identity is satisfied if and only if
\begin{equation}
	\mathbb{J}^{il} \frac{\partial \mathbb{J}^{jk}}{\partial \zeta^l} + \circlearrowleft_{ijk} = 0.
\end{equation}
Denoting the blocks of $\mathbb{J}$ by 
\begin{equation}
	\mathbb{J}_{zz} = \mathbb{M}_p^{-1} \mathbb{J}_c\,,
	\ \ 
	\mathbb{J}_{z\mu} = - \mathbb{J}_{\mu z}^T = \mathbb{J}_c D \mathbb{\Psi}(\bm{Z})\,, 
	\ \  \text{and} \ \ 
	\mathbb{J}_{\mu \mu} = D \mathbb{\Psi}(\bm{Z})^T \mathbb{M}_p \mathbb{J}_c D \mathbb{\Psi}(\bm{Z})\,,
\end{equation}
we find that
\begin{equation}
	\mathbb{J} \frac{\partial \mathbb{J}}{\partial \bm{\zeta}} = 
	\begin{pmatrix}
		\mathbb{J}_{zz} & \mathbb{J}_{z\mu} \\
		\mathbb{J}_{\mu z} & \mathbb{J}_{\mu \mu}
	\end{pmatrix}
	\begin{pmatrix}
		\frac{\partial \mathbb{J}}{\partial \bm{Z}} \\
		0
	\end{pmatrix}
	= \mathbb{J}_{zz} \frac{\partial \mathbb{J}}{\partial \bm{Z}} + \mathbb{J}_{\mu z} \frac{\partial \mathbb{J}}{\partial \bm{Z}}.
\end{equation}
Abusing notation a bit, we write out all four blocks of $\frac{\partial \mathbb{J}^{jk}}{\partial \bm{Z}^l}$ at once for conciseness:
\begin{equation}
	\begin{aligned}
		\mathbb{J}_{zz}^{il} \frac{\partial \mathbb{J}^{jk}}{\partial \bm{Z}^l} 
		&= (\mathbb{M}_p^{-1} \mathbb{J}_c)^{il} 
		\frac{\partial}{\partial Z^l}
		\begin{pmatrix}
				(\mathbb{M}_p^{-1} \mathbb{J}_c)^{jk} & (\mathbb{J}_c D \mathbb{\Psi}(\bm{Z}))^{jk} \\
				(D \mathbb{\Psi}(\bm{Z})^T \mathbb{J}_c)^{jk} & (D \mathbb{\Psi}(\bm{Z})^T \mathbb{M}_p \mathbb{J}_c D \mathbb{\Psi}(\bm{Z}))^{jk}
		\end{pmatrix} \\
		&= 
		(\mathbb{M}_p^{-1})^{i\alpha} \mathbb{J}_c^{\alpha l}
		\begin{pmatrix}
				0^{ljk} & \mathbb{J}_c^{j \beta} \frac{\partial^2 \psi^k}{\partial Z^\beta \partial Z^l} \\
				-\mathbb{J}_c^{k \beta} \frac{\partial^2 \psi^j}{\partial Z^\beta \partial Z^l} & \frac{\partial}{\partial Z^l} \left( \frac{\partial \psi^j}{\partial Z^\beta} \mathbb{M}_p^{\beta \gamma} \mathbb{J}_c^{\gamma \delta} \frac{\partial \psi^k}{\partial Z^\delta} \right)
		\end{pmatrix}.
	\end{aligned}
\end{equation}
When we sum over $(i,j,k)$, one can immediately see that everything except the contributions from the bottom right block vanish since the off diagonal blocks cancel each other. We now consider this bottom right block in detail. Note that $\mathbb{M}_p \mathbb{J}_c = \mathbb{J}_c \mathbb{M}_p$. Hence, 
\begin{equation}
	\begin{aligned}
		\frac{\partial}{\partial Z^l} \left( \frac{\partial \psi^j}{\partial Z^\beta} (\mathbb{M}_p \mathbb{J}_c)^{\beta \delta} \frac{\partial \psi^k}{\partial Z^\delta} \right)
		&= \frac{\partial^2 \psi^j}{\partial Z^l \partial Z^\beta} (\mathbb{M}_p \mathbb{J}_c)^{\beta \delta} \frac{\partial \psi^k}{\partial Z^\delta} + \frac{\partial \psi^j}{\partial Z^\beta} (\mathbb{J}_c \mathbb{M}_p)^{\beta \delta} \frac{\partial \psi^k}{\partial Z^\delta \partial Z^l} \\
		&= \frac{\partial^2 \psi^j}{\partial Z^l \partial Z^\beta} \mathbb{M}_p^{\beta \gamma} \mathbb{J}_c^{\gamma \delta} \frac{\partial \psi^k}{\partial Z^\delta} 
			+ \frac{\partial \psi^j}{\partial Z^\beta} \mathbb{J}_c^{\beta \gamma} \mathbb{M}_p^{\gamma \delta} \frac{\partial \psi^k}{\partial Z^\delta \partial Z^l} \\
		&= \frac{\partial^2 \psi^j}{\partial Z^l \partial Z^\beta} \mathbb{M}_p^{\beta \gamma} \mathbb{J}_c^{\gamma \delta} \frac{\partial \psi^k}{\partial Z^\delta} 
			+ \frac{\partial \psi^j}{\partial Z^\delta} \mathbb{J}_c^{\delta \gamma} \mathbb{M}_p^{\gamma \beta} \frac{\partial \psi^k}{\partial Z^\beta \partial Z^l} \\
		&= \frac{\partial^2 \psi^j}{\partial Z^l \partial Z^\beta} \mathbb{M}_p^{\beta \gamma} \mathbb{J}_c^{\gamma \delta} \frac{\partial \psi^k}{\partial Z^\delta} 
			- \frac{\partial^2 \psi^k}{\partial Z^l \partial Z^\beta} \mathbb{M}_p^{\beta \gamma} \mathbb{J}_c^{\gamma \delta} \frac{\partial \psi^j}{\partial Z^\delta}
	\end{aligned}
	\label{finitejac}
\end{equation}
where, unlike our proof of Jacobi above  for the continuous case that  relied on the theorem of \cite{pjm82}, we retain the second derivative terms and show explicitly they vanish. Also note, in \eqref{finitejac} we arbitrarily reindexed such that $\beta \leftrightarrow \delta$ in the second to last line and used the antisymmetry of $\mathbb{J}_c$. Finally, summing over $(i,j,k)$, we find that these two terms cancel each other. Notice, the cancelation took place even if we only summed over $(j,k)$. Hence, we may replace $\mathbb{J}_{zz} \partial \mathbb{J}/\partial \bm{Z}$ with $\mathbb{J}_{\mu z} \partial \mathbb{J}/\partial \bm{Z}$ and still get cancelation for the same reasons as before. Hence,
\begin{equation}
	\mathbb{J}_{zz}^{il} \frac{\partial \mathbb{J}^{jk}}{\partial \bm{Z}^l} + \circlearrowleft_{i,j,k} = \mathbb{J}_{\mu z}^{il} \frac{\partial \mathbb{J}^{jk}}{\partial \bm{Z}^l} + \circlearrowleft_{i,j,k} = 0.
\end{equation}
Therefore, we find that the Jacobi identity holds for the discrete bracket.